\documentclass[twoside]{article}
\usepackage{fleqn,espcrc2}
\usepackage[dvips]{graphicx}
\usepackage{epsf}
\usepackage{amssymb}
\def\bc{\begin{center}}
\def\ec{\end{center}}
\def\beq{\begin{equation}}
\def\eeq{\end{equation}}
\def\bs{\begin{slide}}
\def\es{\end{slide}}
\newcommand{\bmath}{\begin{displaymath}}
\newcommand{\emath}{\end{displaymath}}
\newcommand{\beqn}{\begin{eqnarray}}
\newcommand{\eeqn}{\end{eqnarray}}
\newcommand{\beqns}{\begin{eqnarray*}}
\newcommand{\eeqns}{\end{eqnarray*}}
\newcommand{\ba}{\begin{array}{c}} 
\newcommand{\bat}{\begin{array}{cc}} 
\newcommand{\ea}{\end{array}}

\newcommand{\Frac}[2]{\frac{\displaystyle #1}{\displaystyle #2}}

\title{Vector form factor of the pion~: A model--independent approach
\thanks{IFIC/02$-$42 and FTUV/02-0919 reports. To appear in the proceedings
of the High--Energy Physics International
Conference on Quantum Chromodynamics, 2--9 July (2002), Montpellier (France).}
}

\author{A. Pich $^{\mbox{\footnotesize a}}$ and J. Portol\'es 
\address{Departament de F\'{\i}sica Te\`orica, IFIC,
CSIC-Universitat de Val\`encia, \\
Edifici d'Instituts d'Investigaci\'o, Apt. Correus 22085, E-46071 Val\`encia,
Spain}}
       
\begin{document}

\begin{abstract}
We study a model--independent parameterization of the vector pion
form factor that arises from the constraints of analyticity and 
unitarity. Our description should be suitable up to 
$\sqrt{s} \simeq 1.2 \, \mbox{GeV}$ and allows a model--independent
determination of the mass of the $\rho(770)$ resonance. We analyse
the experimental data on $\tau^- \rightarrow \pi^- \pi^0 \nu_{\tau}$
and $e^+ e^- \rightarrow \pi^+ \pi^-$
in this framework, and its consequences on the low--energy observables
worked out by chiral perturbation theory. An evaluation of the two pion
contribution to the anomalous magnetic moment of the muon, $a_{\mu}$, and 
to the fine structure constant, $\alpha (M_Z^2)$, is
also performed.
\vspace{1pc}
\end{abstract}

\maketitle

\section{Introduction}

Matrix elements of QCD hadron currents in exclusive processes provide,
from a phenomenological point of view, a detailed knowledge on the
hadronization mechanisms. Their evaluation, however, is a long--standing
problem due to the fact that it involves strong interactions in an
energy region driven by non--perturbative QCD. Within this framework
semileptonic processes, as exclusive hadronic $\tau$ decays 
($\tau^- \rightarrow H^- \nu_{\tau}$) or hadronic cross sections out
of electron--positron annihilation ($e^+ e^- \rightarrow H^0$), furnish
an excellent dynamical system to explore. In the Standard Model their
amplitudes are generically given by
\beq \label{eq:m1}
 M \;  =  \;  C \, {\cal L}^{\mu} \,  {\cal H}_{\mu} \,  , 
\eeq
where $C$ is a factor containing the relevant couplings, 
${\cal L}_{\mu}$ is the leptonic matrix element, easily calculable 
within the theory, and 
\beq \label{eq:m2} 
{\cal H}_{\mu}  \;  = \;    \langle \,  H \,  | \,  J_{\mu} \; 
 e^{i L_{strong}} \, | \, 0 \, \rangle  \, , 
\eeq
with $J_{\mu}$ the vector $V_{\mu}$ ($e^+ e^- \rightarrow H^0$)
or left $V_{\mu} - A_{\mu}$
($\tau^- \rightarrow H^- \nu_{\tau}$) hadron current.
Symmetries help us to define a decomposition of ${\cal H}_{\mu}$ in terms of
the allowed Lorentz structure of implied momenta and a set of 
functions of Lorentz invariants, the {\em form factors} $F_i^H$,
\beq
{\cal H}_{\mu} \; = \; \sum_i  \underbrace{ \; \,  (  \, \ldots \, )_{\mu}^i 
\; \,}_{Lorentz \, struc.}
 F_i^H (q^2, \ldots) \; .
\label{eq:ff}
\eeq
Form factors are the goal of the hadronic matrix elements evaluation and,
as can be noticed from the definition of ${\cal H}_{\mu}$ in Eq.~(\ref{eq:m2}),
are a strong interaction related problem in a non--perturbative regime.
\par
In the last years experiments like ALEPH, CLEO-II, DELPHI, OPAL and
CMD-2 \cite{CMD2,ALEPH97,CLEO00,expe} have provided and important
amount and quality of experimental data on exclusive channels which
phenomenological analysis is now mandatory. However most of these 
analyses are carried out within modelizations (including simplifying
assumptions which may be are not well controlled from QCD itself
\cite{Victoria}) that, while of importance
to get an understanding of the involved dynamics, could give a 
delusive interpretation of data. 
\par
The use of effective actions from QCD
supplies a powerful model--independent procedure to work with. At very
low energies [$E \ll M_{\rho}$, with $M_{\rho}$ the mass of the
$\rho (770)$ resonance] the most important QCD feature is its chiral
symmetry that is realized in chiral perturbation theory ($\chi$PT)
\cite{chipt} with a long and successful set of predictions both in 
strong and electroweak processes \cite{set}. 
At higher energies [$E \sim M_{\rho}$] resonance chiral
theory is the analogous effective theory \cite{rcht}, where the lightest 
resonance fields are kept as explicit degrees of freedom. With the 
addition of dynamical constraints coming from short--distance QCD,
resonance chiral theory becomes a predictive model--independent 
approach. This framework can be combined with S--matrix theory properties.
On general grounds local causality of the interaction translates into
the analyticity properties of amplitudes and, correspondingly, of form
factors.  Being analytic functions in complex variables the behaviour
of form factors at different energy scales is related and, moreover,
they are completely determined by their singularities. Dispersion 
relations embody rigorously these properties and are the appropriate
tool to enforce them.
\par
In this note we recall our work \cite{noi} on the vector form factor
of the pion in the model--independent approach we have just sketched.
We perform a numerical analysis of the recent $e^+ e^-$ CMD-2 data 
\cite{CMD2} and we reanalyse the $\tau$ ALEPH data \cite{ALEPH97} when
corrected by isospin breaking effects \cite{Vincenzo}. The output of
these analyses is a determination of the $\rho (770)^{\pm,0}$ masses,
the low--energy parameters of the vector pion form factor, data on
the $\omega(782)$ resonance, particularly the $\rho-\omega$ mixing,
and a new evaluation of the two--pion contribution to the anomalous 
magnetic moment of the muon $a_{\mu}$ and to $\Delta \alpha (M_Z^2)$.

\section{Vector form factor of the pion}

The pion vector form factor, $F_V(s)$ is defined through
\beq
\langle \, \pi^+(p') \, \pi^-(p) \, | \, V_{\mu}^3 \, | \, 0 \, \rangle
\, = \, (p-p')_{\mu} \, F_V(s) \; , 
\label{eq:pff}
\eeq
where $s=q^2=(p+p')^2$ and $V_{\mu}^3$ is the third component of the 
vector current associated to the $SU(3)$ flavour symmetry of the QCD
Lagrangian. This form factor drives the isovector hadronic part of 
$e^+ e^- \rightarrow \pi^+ \pi^-$ and, in the isospin limit, of
$\tau^- \rightarrow \pi^- \pi^0 \nu_{\tau}$.
At very low energies, $F_V(s)$ has been
studied in the $\chi$PT framework up to ${\cal O}(p^6)$ \cite{fv1,fv2}. A
successful study at the $\rho(770)$ energy scale has been carried out
in the resonance chiral theory in Ref. \cite{pg97}.
\par
Analyticity and unitarity properties of $F_V(s)$ tightly constrain, on general
grounds, the structure of the form factor \cite{noi,pg97}. Elastic unitarity
and Watson final--state theorem relate the imaginary part of $F_V(s)$ to 
the partial wave amplitude $t_1^1$ for $\pi \pi$ elastic scattering, with
angular momentum and isospin equal to one, as
\beq
\mbox{Im} \, F_V(s+i\varepsilon) \, = \; e^{i \delta_1^1} \, \sin (\delta_1^1) 
\, F_V(s)^* \, , 
\eeq
that shows that the phase of $F_V(s)$ must be $\delta_1^1$. Thus analyticity
and unitarity properties of the form factor are accomplished by demanding
that it should satisfy a n--subtracted dispersion relation with the 
Omn\`es solution \cite{pg97}
\beqn
F_V(s)  & = &  \exp \left\{ \sum_{k=0}^{n-1} \Frac{s^k}{k!} 
\Frac{d^k}{ds^k} \ln F_V(s)|_{s=0} \; \right. \nonumber \\
& & \; \; \; \; \; \; \; \left. + \, \Frac{s^n}{\pi}
 \int_{4 m_{\pi}^2}^{\infty} 
\Frac{dz}{z^n} \Frac{\delta_1^1 (z)}{z - s - i \varepsilon} \right\} .
\label{eq:solom}
\eeqn
This solution is strictly valid only below the inelastic threshold 
($s  <  16 m_{\pi}^2$), however higher multiplicity intermediate states
are suppressed by phase space and ordinary chiral counting. The
$\delta_1^1(s)$ phase--shift, in Eq.~(\ref{eq:solom}), is
rather well known, experimentally, up to $E \sim 2 \, \mbox{GeV}$
\cite{ochs}. 
\par
With an appropriate number of subtractions we can parameterize
$F_V(s)$ with the subtraction constants appearing in the first term of the
exponential in Eq.~(\ref{eq:solom}). In Ref.~\cite{noi} we have used
three subtractions :
\beqn
F_V(s)  & = &  \exp \left\{ \alpha_1 s \, + \, \Frac{1}{2} \alpha_2
s^2 \, \right. \nonumber \\
& & \; \; \; \; \; \; + \left. \Frac{s^3}{\pi} \int_{4 m_{\pi}^2}^{\Lambda^2}
\Frac{dz}{z^3} \Frac{\delta_1^1(z)}{z - s - i \varepsilon} \, \right\} ,
\label{eq:sol3}
\eeqn
\noindent
where we have introduced an upper cut in the integration, $\Lambda$. 
This cut--off has to be taken high enough not to spoil the, a priori, 
infinite interval of integration, but low enough that the integrand  is
well known in the interval. The two subtraction constants $\alpha_1$ and
$\alpha_2$ (a third one is fixed by the normalization $F_V(0) = 1$) are
related with the squared charge radius of the pion 
$\langle r^2 \rangle_V^{\pi}$ and the quadratic term $c_V^{\pi}$ in the
low--energy expansion 
\begin{table}
\begin{center}
\begin{tabular}{|c|c|c|} 
\hline
\multicolumn{1}{|c|}{Source} &
\multicolumn{1}{|c|}{$M_{\rho^{\pm}} (\mbox{MeV})$} &
\multicolumn{1}{|c|}{$M_{\rho^0} (\mbox{MeV})$} \\
\hline
\hline
Our fit &  $775.9 \pm 0.5$ &  $777.8 \pm 0.7$ \\
\hline
Ref.~\protect{\cite{TY01}} & $773.8 \pm 0.6$ & $ 772.6 \pm 0.5$ \\
\hline
Average~\protect{\cite{RPP02}} & \multicolumn{2}{|c|}{$775.9 \pm 0.5$} \\
\hline
\multicolumn{3}{c}{}
\end{tabular} 
\caption{Comparison of our results for $M_{\rho^{\pm}}$ and 
$M_{\rho^0}$ with other recent figures. The average value 
\protect{\cite{RPP02}} corresponds to $e^+ e^-$ and $\tau$ data
analyses only.}
\end{center}
\vspace*{-0.4cm}
\end{table}
\begin{figure}[htb]
\vspace*{-0.9cm}
\hspace*{-0.8cm} 
\includegraphics*[angle=-90,scale=0.33,clip]{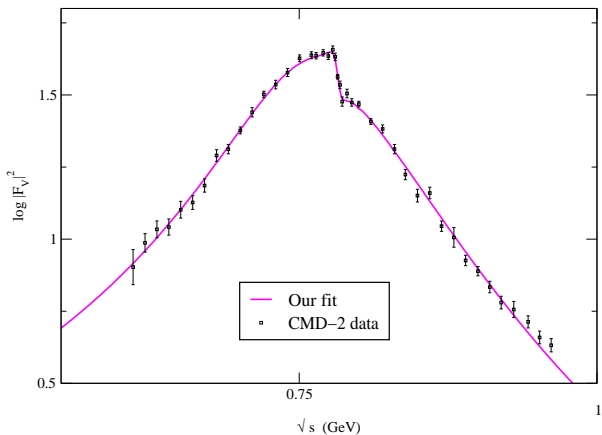}
\vspace*{-1.5cm} 
\caption{\label{fig:f1} Comparison of experimental data from
$e^+ e^- \rightarrow \pi^+ \pi^-$ by CMD-2 \cite{CMD2} with the
results of our fit.}
\vspace*{-0.4cm}
\end{figure}
\beq
F_V(s) \, = \, 1 + \Frac{1}{6} \langle r^2 \rangle_V^{\pi} s \, + \,
c_V^{\pi} s^2 \, + \, {\cal O}(s^3) .
\label{eq:chirex}
\eeq
The input
of the $\delta_1^1(s)$ phase--shift is included as follows. 
Resonance chiral theory and vector meson dominance provide a 
model--independent analytic expression that describes properly the
$\rho(770)$ contribution \cite{pg97}
\beq
\delta_1^1(s) \, = \, \arctan \left\{ \Frac{M_{\rho}
\Gamma_{\rho}(s)}{M_{\rho}^2 - s} \right\} \, ,
\label{eq:d11r}
\eeq
with $\Gamma_{\rho}(s)$ the off--shell $\rho(770)$ width as computed
in Ref.~\cite{noi1}. This phase--shift is accurate up to 
$E \sim 1 \, \mbox{GeV}$. At higher energies heavier resonances with the
same quantum numbers pop up and we use the available experimental data
from Ochs \cite{ochs}.

\begin{table}[h]
\vspace*{-0.1cm}
\begin{center}
\begin{tabular}{|c|c|c|} 
\hline
\multicolumn{1}{|c|}{Source} &
\multicolumn{1}{|c|}{$\langle r^2 \rangle_V^{\pi} (\mbox{GeV}^{-2})$} &
\multicolumn{1}{|c|}{$c_V^{\pi} (\mbox{GeV}^{-4})$} \\
\hline
\hline
Our fit ($\tau$) &  $11.0 \pm 0.3$ &  $3.84 \pm 0.03$ \\
\hline
Our fit ($e^+ e^-$) & $11.5 \pm 0.2$ & $3.73 \pm 0.02$ \\
\hline
${\cal O}(p^6) \, \chi$PT \protect{\cite{fv2}} & $11.22 \pm 0.41$
& $3.85 \pm 0.60$  \\
\hline
Ref.~\protect{\cite{TY01}}  & $11.17 \pm 0.05$ 
& $ 3.60 \pm 0.03$ \\
\hline
\multicolumn{3}{c}{}
\end{tabular} 
\caption{Comparison of our results for the low--energy parameters
of the pion form factor with other recent figures.}
\end{center}
\vspace*{-0.3cm}
\end{table}
\begin{figure}[h]
\vspace*{-1cm}
\hspace*{-0.7cm} 
\includegraphics*[angle=-90,scale=0.32,clip]{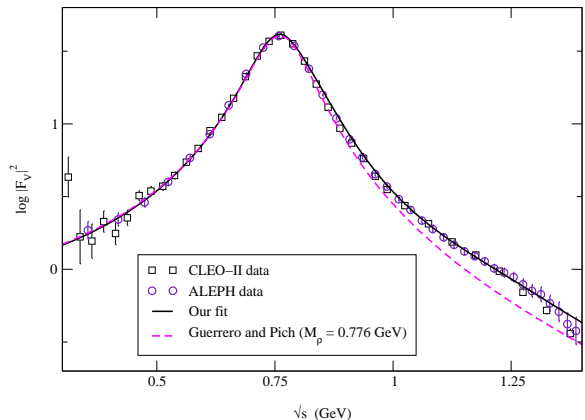}
\vspace*{-1.5cm} 
\caption{\label{fig:f2} Comparison of experimental data from 
$\tau^- \rightarrow \pi^- \pi^0 \nu_{\tau}$ decays by 
ALEPH \cite{ALEPH97} and CLEO-II \cite{CLEO00} with the 
prediction of Ref.~\cite{pg97} and our fit to the ALEPH data.}
\vspace*{-0.5cm}
\end{figure}
\begin{table}[h]
\vspace*{-0.1cm}
\begin{center}
\begin{tabular}{|c|c|c|} 
\hline
\multicolumn{1}{|c|}{$\sqrt{s}_{max} (\mbox{GeV})$} &
\multicolumn{1}{|c|}{$a_{\mu}^{\pi \pi} |_{\tau} \times 10^{10}$} &
\multicolumn{1}{|c|}{$a_{\mu}^{\pi \pi} |_{e^+ e^-} \times 10^{10}$} \\
\hline
\hline
0.5 &  $55.9 \pm 0.5$ &  $56.7 \pm 0.6$ \\
\hline
0.9 & $488 \pm 7$ & $475 \pm 5$ \\
\hline
1.0  & $507 \pm 7$ & $490 \pm 6$ \\
\hline
1.1 & $513 \pm 8$ & $494 \pm 6$  \\
\hline
\multicolumn{3}{c}{}
\end{tabular} 
\caption{Results for the two--pion contribution to $a_{\mu}$ from 
the analyses of $\tau$ and $e^+ e^-$ data and for different values of
the $\sqrt{s}_{max}$ cut--off.}
\end{center}
\vspace*{-0.8cm}
\end{table}

\begin{table}
\begin{center}
\begin{tabular}{|c|c|c|} 
\hline
\multicolumn{1}{|c|}{$\sqrt{s}_{max} (\mbox{GeV})$} &
\multicolumn{1}{|c|}{$10^{4} \,\Delta \alpha (M_Z^2)|_{\tau}$} &
\multicolumn{1}{|c|}{$10^{4} \, \Delta \alpha (M_Z^2)|_{e^+ e^-}$} \\
\hline
\hline
0.9 & $31.9 \pm 0.5$ & $30.7  \pm 0.4$ \\
\hline
1.0  & $34.0 \pm 0.5$ & $32.4 \pm 0.4$ \\
\hline
1.1 & $34.8 \pm 0.6$ & $33.0 \pm 0.5$  \\
\hline
\multicolumn{3}{c}{}
\end{tabular} 
\caption{Results for the two--pion contribution to 
$\Delta \alpha (M_Z^2)$ from 
the analyses of $\tau$ and $e^+ e^-$ data and for different values of
the $\sqrt{s}_{max}$ cut--off.}
\end{center}
\vspace*{-1cm}
\end{table}

\par
$F_V(s)$ endows the hadronic dynamics in the $e^+ e^- \rightarrow \pi^+ \pi^-$
process and, in the isospin limit in the
$\tau^- \rightarrow \pi^- \pi^0 \nu_{\tau}$ decay. In Ref.~\cite{noi}
we analysed the $\tau$ ALEPH \cite{ALEPH97} data where radiative
corrections were not taken into account. In this note we reanalyse these
data when corrected for isospin breaking effects recently
computed \cite{Vincenzo}. Analogously we study the recent experimental
data on $e^+ e^- \rightarrow \pi^+ \pi^-$ en the $\rho(770)$ energy
region by CMD-2 \cite{CMD2}. To analyse the $e^+ e^-$ data we need to
include the effect of the $\omega(782) \rightarrow \pi^+ \pi^- $ process
in our form factor.
This we do through a $\rho-\omega$ mixing term defined as in 
Ref.~\cite{Urech}.
\par
The results of our fit to CMD-2 data, with $\chi^2 /dof = 45.2/37$,
are shown in Fig.~\ref{fig:f1}
while the $M_{\rho^0}$ mass and low--energy parameters are given in 
Tables~1~and~2. In addition we get 
$M_{\omega} = (781.8 \pm 0.3) \, \mbox{MeV}$, 
$\Gamma_{\omega} = (9.3 \pm 1.6) \, \mbox{MeV}$ and 
$\Theta_{\rho \omega} = (-3.3 \pm 0.5) \times 10^{-3}\, \mbox{GeV}^2$. 
The analysis of the $\tau$ ALEPH data gives the results shown in 
Fig.~\ref{fig:f2}. The fit has a $\chi^2 /dof = 30.2 /21$ and
the values of $M_{\rho^{\pm}}$ and the low--energy parameters can also be
read in Tables~1~and~2.
We also get $\Delta M_{\rho^{\pm}-\rho^0} = (-1.9 \pm 0.9) \, \mbox{MeV}$
and $\Delta \Gamma_{\rho^{\pm}-\rho^0} = (-0.2 \pm 0.6) \, \mbox{MeV}$,
to be compared with the figures in Ref.~\cite{RPP02}~: 
$\Delta M_{\rho^{\pm}-\rho^0} = (-0.4 \pm 0.8) \, \mbox{MeV}$
and $\Delta \Gamma_{\rho^{\pm}-\rho^0} = (0.1 \pm 1.9) \, \mbox{MeV}$.
\par
Finally, in Tables~3 and~4 we show the results for the two--pion vacuum
polarization contribution
to the anomalous magnetic moment of the muon, $a_{\mu}^{\pi \pi}$ and to
the shift in the fine--structure constant $\Delta \alpha (M_Z^2)^{\pi \pi}$,
for different values of $\sqrt{s}_{max}$ (the upper limit of the hadronic
invariant mass in the dispersion integral that provides the hadron vacuum
polarization contribution to both observables)  
and for the form factors
coming from the analyses of $e^+ e^-$ and $\tau$ data. Our results 
compare well with the recent computation in Ref.~\cite{Davier}. 
\vspace*{0.2cm} \\
\noindent
{\bf Acknowledgements}  \\
We wish to thank S.~Narison and his team for the great organization
of the QCD02 Conference. We also thank V. Cirigliano for his help with 
the isospin breaking corrections. This work has been supported in part by 
TMR, EC Contract No. ERB FMRX-CT98-0169, by MCYT (Spain) under grant
FPA-2001-3031 and by ERDF funds from the EU Commission.

\end{document}